\documentclass[sigconf]{acmart}
\usepackage{tabularx}
\usepackage{multirow}

\usepackage{booktabs}
\usepackage{stfloats}
\usepackage{subfigure}
\usepackage{hyperref}

\AtBeginDocument{
  \providecommand\BibTeX{{
    \normalfont B\kern-0.5em{\scshape i\kern-0.25em b}\kern-0.8em\TeX}}}

\copyrightyear{2021} 
\acmYear{2021} 
\setcopyright{acmcopyright}
\acmConference[MMAsia '21]{ACM Multimedia Asia}{December 1--3, 2021}{Gold Coast, Australia}
\acmBooktitle{ACM Multimedia Asia (MMAsia '21), December 1--3, 2021, Gold Coast, Australia}
\acmPrice{15.00}
\acmDOI{10.1145/3469877.3490595}
\acmISBN{978-1-4503-8607-4/21/12}

\begin{document}

\title{Hierarchical Deep Residual Reasoning\\ for Temporal Moment Localization}

\author{Ziyang Ma, Xianjing Han, Xuemeng Song\textsuperscript{*}, Yiran Cui, Liqiang Nie}\thanks{*Xuemeng Song is the corresponding author}
\affiliation
{\institution{Shandong University, Shandong, China}
 \country{}}
\email{ziyang.ma@mail.sdu.edu.cn, {hanxianjing2018, sxmustc, conyrol120, nieliqiang}@gmail.com}

\renewcommand{\shortauthors}{Z. Ma~\textit{et al.}}

\begin{abstract}
Temporal Moment Localization (TML) in untrimmed videos is a challenging task in the field of multimedia, which aims at localizing the start and end points of the activity in the video, described by a sentence query. 
Existing methods mainly focus on mining the correlation between video and sentence representations or investigating the fusion manner of the two modalities. 
These works mainly understand the video and sentence coarsely, ignoring the fact that a sentence can be understood from various semantics, and the dominant words affecting the moment localization in the semantics are the action and object reference. 
Toward this end, we propose a Hierarchical Deep Residual Reasoning (HDRR) model, which decomposes the video and sentence into multi-level representations with different semantics to achieve a finer-grained localization. 
Furthermore, considering that videos with different resolution and sentences with different length have different difficulty in understanding, 
we design the simple yet effective Res-BiGRUs for feature fusion, which is able to grasp the useful information in a self-adapting manner. 
Extensive experiments conducted on Charades-STA and ActivityNet-Captions datasets demonstrate the superiority of our HDRR model compared with other state-of-the-art methods. 
\end{abstract}


\maketitle

\begin{figure}[t]
  \centering
  \includegraphics[width=0.9\linewidth]{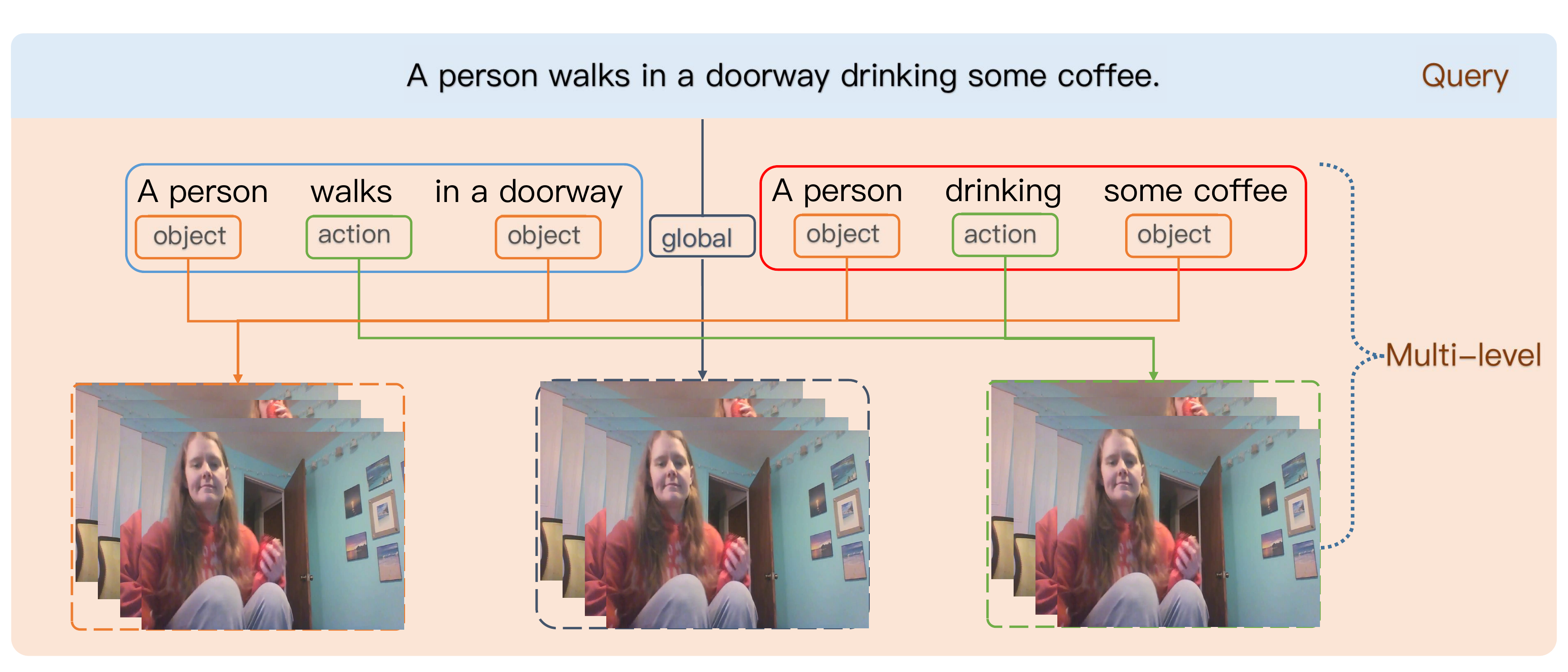}
  \vspace{-0.4cm}
  \caption{Illustration of the multi-level representations of the sentence and video.}
  \label{head-figure}
  \vspace{-0.6cm}
\end{figure}

\section{Introduction}
The explosive increasing of the video data in our daily life makes the processing and automatic understanding to the video significantly necessary. 
Due to its huge application requirements in fields of security (e.g. surveillance) and network media (e.g. fragment retrieval), the task of Temporal Moment Localization (TML) \cite{liu2018cross, cao2020strong, qu2020fine, liu2020jointly} that aims at localizing the start and end points of the activity described by the sentence query in an untrimmed video, attracts the attention of plenty of researchers. 
It is worth noting that compared with traditional video understanding tasks, such as the action detection that detects actions from a limited pre-defined set \cite{richard2016temporal, zhong2018step, yang2019step}, TML is more flexible, since it supports the natural language based query. Accordingly, the task of TML is more challenging. 

Although existing studies have achieved promising progress, they mainly suffer from the following two limitations: 1)  \textit{lack the explicit semantic-oriented representation learning}. Since the representation of the sentence plays a pivotal role in TML, several efforts \cite{chen2018temporally, jiang2019cross, zhang2019cross, chen2019semantic, xu2019multilevel, ge2019mac} have investigated the fine-grained (e.g., word-level and phrase-level) representations of the sentence, for localizing the target 
moments in video. Nevertheless, they neglect the fact that the key words in the sentence query referring to the target moment are usually actions and objects references. In other words, they overlook the potential of the actions and objects in the sentence in localizing the target moment. 
And 2) \textit{lack the adaptive multi-modal fusion}. As two modalities are involved in the task of TML, one essential problem of TML is how to effectively fuse the representations of the two modalities to facilitate the localization of the target moment. Although existing methods have achieved great success with either the graph neural network \cite{zhang2019man, zhou2020graph, liu2020jointly, zhang2021multi} or the iterative attention \cite{yuan2019find, qu2020fine, liu2020jointly}, they neglect that the information embedded in different videos and sentences may have different levels of difficulty to be understood. 
Intuitively, shorter and high-quality videos are easier to be analysed, compared with the longer and low-quality ones. Analogously, the longer and complex sentences are also more difficult to be reasoned compared with the shorter ones. For example, the sentence query ``A man begins changing as he goes upstairs'' is harder to be understood by the model than ``A man is eating''. Therefore, it is necessary to devise an adaptive fusion scheme to handle the real-world application scenario, where both simple and complex samples are possible as the input.

\begin{figure*}[t]
  \centering
  \includegraphics[width=0.9\linewidth]{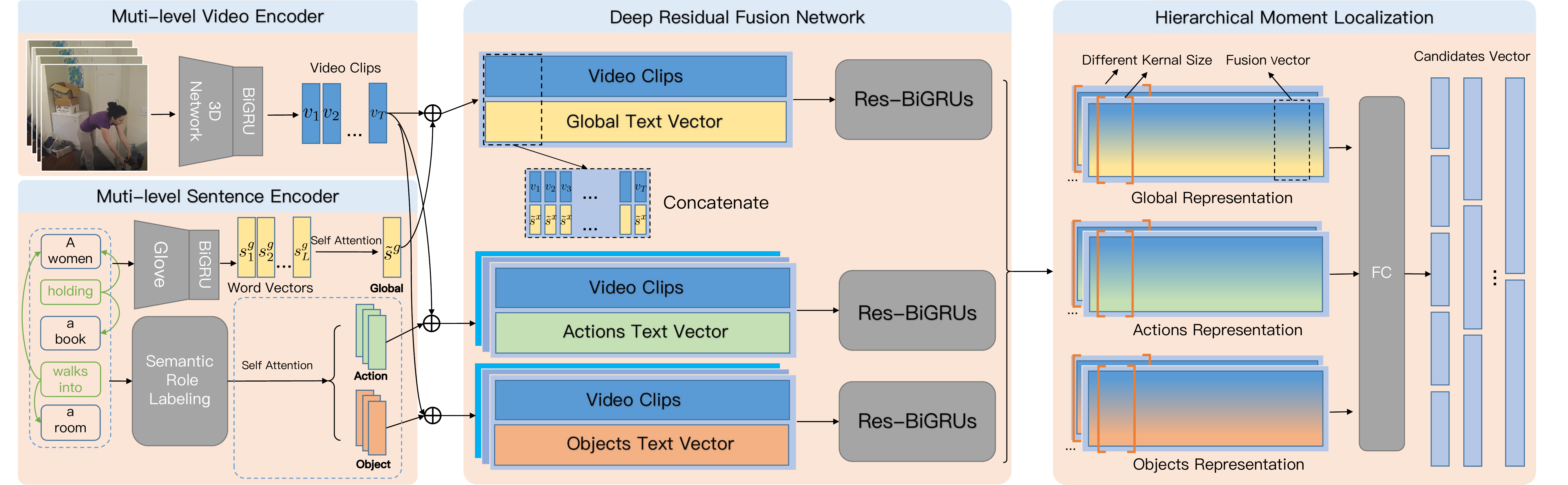}
  \vspace{-0.4cm}
  \caption{Pipeline of the proposed Hierarchical Deep Residual Reasoning. Firstly, the video and the sentence are encoded in multi-level layers, then multi-layer matching is carried out, and our Res-BiGRU module is used for self-adapting fusion. Finally, our hierarchical temporal localization is obtained through hierarchical convolution. }
  \Description{Pipeline of the proposed Hierarchical Deep Residual Reasoning. }
  \label{model-pipeline}
  \vspace{-0.4cm}
\end{figure*}

Toward address the above limitations, as shown in Figure~\ref{head-figure}, we propose to explore the multi-level representations of the video and sentence, including not only the coarse global representation, but also the fine-grained semantic-oriented (action- and object-oriented) representations. Concretely, we introduce a Hierarchical Deep Residual Reasoning (HDRR) model to tackle the task of TML, as shown in Figure~\ref{model-pipeline}, where we represent the video and sentence at different levels to thoroughly capture the correlation between video clips and the sentence query, and design the Res-BiGRUs to boost the adaptive fusion of the two modalities. In particular, we employ the pre-trained BERT to perform Semantic Role Labeling (SRL) to the sentence and get the global, action, and object level representations. 
To learn the matching degree between the video clips and the sentence query, we devise the Res-BiGRUs to adaptively fuse the two modalities and avoid the information loss caused by the representation smoothing. Based on the fused multi-level representations, we hierarchically localize the video moment according to the sentence query.The main contributions of our work are threefold:
\begin{itemize}
  \item We propose a Hierarchical Deep Residual Reasoning (HDRR) model, which decomposes the video and sentence into multi-level representations with different semantics to strengthen the temporal moment localization with hierarchical reasoning. 
  \item Considering that different videos and sentences may have different levels of difficulty to be reasoned, we design the Res-BiGRUs for the self-adaptive feature fusion, which also avoids the information loss caused by the representation smoothing as a byproduct.
  \item Extensive experiments on the Charades-STA and ActivityNet-Captions datasets demonstrate the superiority of our model over other state-of-the-art methods. We will release the code\footnote{\url{https://github.com/ddlBoJack/HDRR}} to benefit other researchers. 
\end{itemize}

\section{Related Work}
\label{Related Work}
\subsection{Temporal Moment Localization}
The task of TML in untrimmed videos was first introduced by Gao~\textit{et al.}~\cite{gao2017tall} and Hendricks~\textit{et al.}\cite{anne2017localizing}.
Early works on TML mainly focus on mining the correlation between the video clip and sentence representations~\cite{gao2017tall, anne2017localizing, liu2018cross, liu2018attentive, yuan2019find}. 
For example, Liu~\textit{et al.}~\cite{liu2018attentive} enhanced the sentence representation by combining contextual video features with the attention mechanism.
Although early methods obtain compelling progress, they overlook the fine-grained interaction between the video and the sentence. To address this issue, some later methods \cite{chen2018temporally, jiang2019cross, zhang2019cross, chen2019semantic, xu2019multilevel, ge2019mac} shed light on the fine-grained understanding of the two modalities. 
For example, Ge~\textit{et al.}~\cite{ge2019mac} added predefined actions feature in the process of matching sentence and video. One key limitation of these methods is that the predefined actions are limited and can hardly cover all kinds of actions occurred in testing phase. Besides, they ignored the objects information, which also plays a pivotal role in localizing the target moment.
Recently, Mun~\textit{et al.}~\cite{mun2020local} considered the semantic entities such as actions and objects in the sentence, but they entangled the interaction between different semantic entities with the video, which may require more search space to capture fine-grained localization.
Beyond that, in our work, we propose the multi-grained moment localization, where the video and sentence is represented with multi-levels, i.e., the global level and the fine-grained action/object level. 

\subsection{Semantic Role Labeling}
Semantic Role Labeling (SRL) raised by Charles J. Fillmore~\cite{fillmore1976frame} is the task of determining the latent predicate argument structure of a sentence and hence facilitate the question answering, like who did what to whom.
Traditional SRL systems~\cite{carreras2005introduction} are mainly based on the syntactic analysis. However, complete syntactic analysis requires all the syntactic information contained in the sentence, and minor errors in syntactic analysis may lead to severe errors in the sentence reasoning. 
The block-based SRL approach solves SRL as a sequence tagging problem, which is usually solved with BERT. Simple BERT\cite{Shi2019SimpleBM} is an implementation of the BERT based model, which is currently the state-of-the-art single model for English PropBank SRL~\cite{palmer2005proposition}. 
Inspired by the success of SRL in sentence structure analysis, in this paper, we employ the BERT-based SRL model~\cite{Shi2019SimpleBM} to extract the actions and objects in the sentence query and hence facilitate the hierarchical moment localization. 

\section{Method}
\label{Method}
In this section, we first give the problem formulation, then detailed the proposed Hierarchical Deep Residual Reasoning model.

\subsection{Problem Formulation}
Formally, suppose we have an untrimmed video $V$ and a sentence query $S$. Owing to that the target moment usually involves multiple continuous units of the video, we split the untrimmed video and represent it with $V=\{v_t\}_{t=1}^T$, where $v_t$ is the $t$-th unit of the video and $T$ denotes the total number of the units. 
The sentence query is represented as $S=\left\{{s}_{l}\right\}_{l=1}^{L}$, where ${s}_{l}$ is the $l$-th word in the sentence and $L$ is the length of the sentence. Overall, the task is to determine the start and end points $({\xi }^{s},{\xi }^{e})$ of the target moment in the video corresponding to the sentence query. 

\subsection{Hierarchical Deep Residual Reasoning}
\subsubsection{Multi-level Sentence Encoder}
We first present the multi-level sentence encoder used for representing the sentence query, from not only the coarse-grained global level, but also fine-grained action and object levels. 

\indent \textbf{Global Level Feature Extraction.}
Regarding the global-level encoding of the sentence, similar with previous studies~\cite{liu2018cross,yuan2019find,liu2020jointly, qu2020fine}, we use GloVe~\cite{pennington2014glove} to obtain the word embedding $\mathbf{s}_{l}$ for each word in the sentence.
Then we use BiGRU to encode the the contextual information among the words as follows,
\begin{equation}
  \left\{
  \label{eq1}
  \begin{aligned}
  \overrightarrow{\mathbf{h}}_{l}^{s} &=\overrightarrow{GRU}^{s}\left(\mathbf{s}_{l}, \overrightarrow{\mathbf{h}}_{l-1}^{s}\right), \\
  \overleftarrow{\mathbf{h}}_{l}^{s} &=\overleftarrow{GRU}^{s}\left(\mathbf{s}_{l}, \overleftarrow{\mathbf{h}}_{l+1}^{s}\right), \\
  \mathbf{s}_{l}^{g} &=f^{s}\left(\overrightarrow{\mathbf{h}}_{l}^{s} \| \overleftarrow{\mathbf{h}}_{l}^{s}\right),
  \end{aligned}
  \right.
\end{equation}
where $\overrightarrow{\mathbf{h}}_l^{s}$ and $\overleftarrow{\mathbf{h}}_l^{s}$ denote the $l$-th hidden state vectors generated by the forward and the backward GRU, respectively. $\|$ denotes the concatenation operation and $f^s$ denotes the fully-connected layer followed with ReLU activation function to fuse the hidden states for each word derived by the two GRUs. $\mathbf{s}_{l}^{g}\in \mathbb{R}^{d_s}$ refers to the latent representation of the $g$-th word, and $d_s$ is the dimension of the representation.
Finally, the global representation of the sentence is denoted as $\mathbf{S}^g=[\mathbf{s}_{1}^{g},\mathbf{s}_{2}^{g},\cdots,\mathbf{s}_{L}^{g}], \mathbf{S}^g \in \mathbb{R}^{L \times d_s}$.

\indent \textbf{Semantic Level Feature Extraction.}
To explore the fine-grained information of sentences, similar to \cite{chen2020fine}, we use the pre-trained SRL BERT \cite{Shi2019SimpleBM} to obtain the semantic role of each word in the sentence, which first decomposes the sentence into multiple frames and then analyses the semantic role of words in each frame.
For example, with the SRL BERT, the sentence ``A woman holding a book walks into a room'' can yield two frames: ``a woman (object) / holding (action) / a book (object)'' and ``a woman (object) / walks into (action) / a room (object)'', where the objects ``a woman'' and ``a book'' are correlated by the action ``holding'', and the objects ``a woman'' and ``a room'' are associated with the action ``walks into'', respectively. 

Formally, we use the SRL BERT to get the semantic role for the given sentence $S$ and derive the object and action level representations denoted as $\mathbf{S}^o=\{\mathbf{s}^o_1, \mathbf{s}^o_2, \cdots, \mathbf{s}^o_{L_o}\}$ and $\mathbf{S}^a=\{\mathbf{s}^a_1, \mathbf{s}^a_2, \cdots, \mathbf{s}^a_{L_a}\}$, where $L_a$ and $L_o$ are the total number of the words with action and object role, respectively. Note that to ease the calculation, we pad the non-action and non-object words in the sentence with $0$s to regularize the dimension of $\mathbf{S}^o$ and $\mathbf{S}^a$ to $\mathbb{R}^{L \times d_s}$, respectively. 

\indent \textbf{Attentive Representation Learning. }
To highlight the prominent information of the representation at each level, we further employ the multi-head self-attention~\cite{vaswani2017attention} that has turned out to be effective in capturing the the useful contextual information, to derive the final sentence representations as follows,
\begin{equation}
  \left\{
  \begin{aligned}
  \operatorname{h}_{i}&=\operatorname{Softmax}\left(\frac{\mathbf{S}^x \mathbf{W}_{i}^{Q,x}\left(\mathbf{S}^x \mathbf{W}_{i}^{K,x}\right)^{\top}}{\sqrt{d_{h}}}\right) \mathbf{S}^x \mathbf{W}_{i}^{V,x},\\
  {\tilde{\mathbf{s}}^x}&=\mathbf{W}^{x}_{O} \text {Concat}\left({h}_{1}, \ldots, \text{h}_{n}\right),\\
  \end{aligned}
  \right.
\end{equation}
where $\mathbf{S}^x \in \mathbb{R}^{L \times d_s}, x \in \{g, a, o\}$ represents the representations of the sentence at different levels, including the global, and semantic levels. $\mathbf{W}_{i}^{Q,x}, \mathbf{W}_{i}^{K,x}, \mathbf{W}_{i}^{V,x} \in \mathbb{R}^{d_{s} \times d_{h}}$ are the learnable parameters of the $i$-th head. $d_{s}=d_{h}\times n$, where $d_h$ is the size of the output feature for each head and $n$ is the number of the parallel heads. The output of each head $h_i$ is concatenated and projected by a linear transformation $\mathbf{W}^{x}_{O} \in \mathbb{R}^{1 \times L}$ to generate the final different level sentence representations $\tilde{\mathbf{s}}^x \in \mathbb{R}^{1 \times d_s}, x\in\{g, a, o\}$. 

\subsubsection{Multi-level Video Encoder}\label{Hierarchical Video Encoder}
To facilitate the correspondence modeling between the video clip and the sentence, we also characterize the video at different levels, i.e., the global, action, and object levels. 

\textbf{Global Level Encoding.}
We first use the pre-trained 3D network~\cite{tran2015learning, carreira2017quo, wang2016temporal} to extract the visual feature for the video, denoted as $\mathbf{V}=[\mathbf{v}_{1},\mathbf{v}_{2},\cdots,\mathbf{v}_{T}]$, where $\mathbf{v}_{t}$ is the visual feature of the $t$-th unit.
Similar with the global level encoding of the sentence, we also use BiGRU to excavate the contextual information among the video unit sequence, which can be formulated as follows, 
  \begin{equation}
    \begin{aligned}
    \mathbf{V}^{g}=BiGRU_v(\mathbf{V})
    \end{aligned}
  \end{equation}
where $BiGRU$ denotes the BiGRU network following the same architecture of Equation~\ref{eq1}.
$\mathbf{V}^g=[\mathbf{v}_{1}^{g},\mathbf{v}_{2}^{g},\cdots,\mathbf{v}_{T}^{g}]$ represents the obtained global level representation of the video, where $\mathbf{v}_{t}^{g}$ stands for the global level representation of the $t$-th video unit.

\textbf{Semantic Level Encoding.} Apparently, the semantic level representations of video can be hardly explicitly extracted like that of the sentence. Thus, we employ two fully-connected layers to highlight the action and object information in the video, respectively. Formally, we have, 
\begin{equation}
 \mathbf{v}_t^x = f^x(\mathbf{v}_t^g), x \in \{a,o\}
\end{equation}
where $f^x, x \in \{a,o\}$ is a fully-connected layer, in which $a$ represents for ``action'' and $o$ represents for ``object''. $\mathbf{v}_t^x, x \in \{a,o\}$ refers to the action- and object-oriented representations of the $t$-th video unit. Ultimately, the action- and object-oriented representations of the whole video can be denoted as $\mathbf{V}^a=[\mathbf{v}_{1}^{a},\mathbf{v}_{2}^{a},\cdots,\mathbf{v}_{T}^{a}]$ and $\mathbf{V}^o=[\mathbf{v}_{1}^{o},\mathbf{v}_{2}^{o},\cdots,\mathbf{v}_{T}^{o}]$, respectively. 

\subsubsection{Deep Residual Fusion Network}
\label{Deep Residual Fusion Network}

\begin{figure}[!t]
 \centering
 \includegraphics[width=1\linewidth]{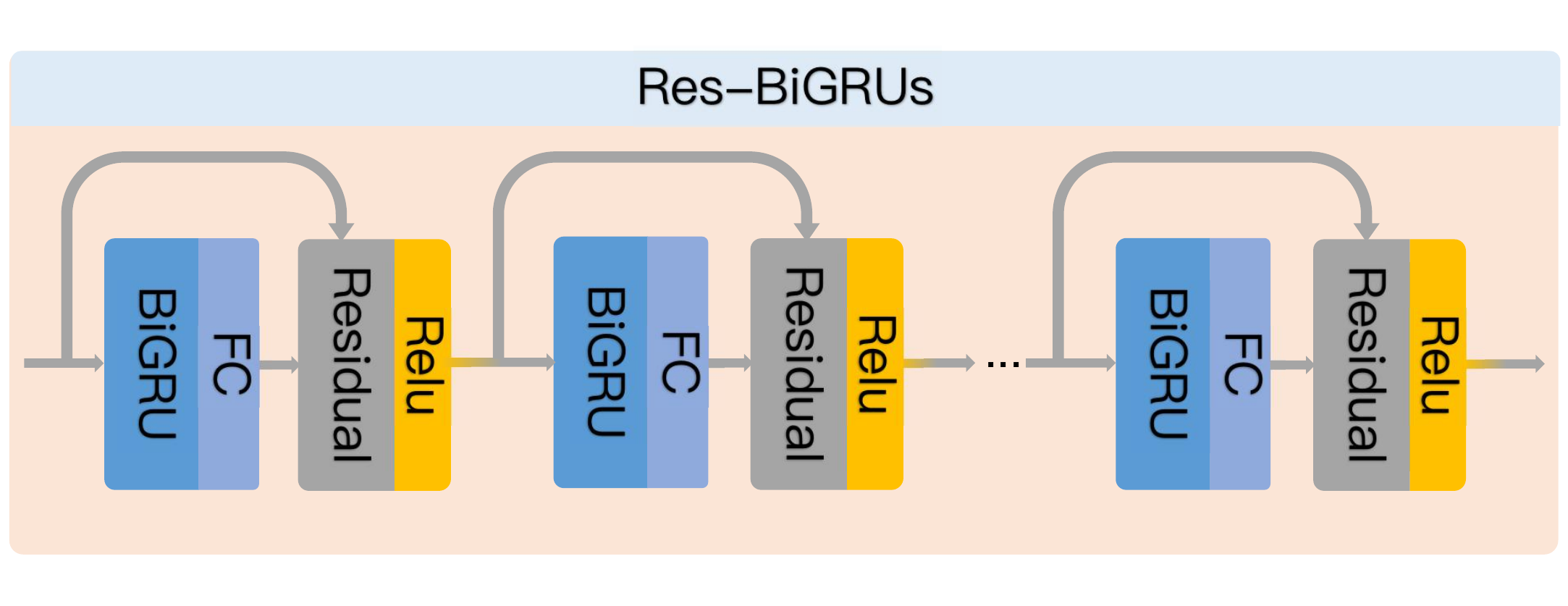}
 \vspace{-0.4cm}
 \caption{The illustration of Res-BiGRUs.}
 \vspace{-0.4cm}
 \label{Res-BiGRU}
\end{figure}

In order to adaptively fuse the representations of the video and sentence at each level, we devise the Res-BiGRUs based fusion network, which is the stacking of multiple BiGRU layer, as shown in Figure \ref{Res-BiGRU}. The deep structure of the Res-BiGRUs ensures the sufficient exploitation to the difficult samples, while the residual design is employed to automatically maintain the characteristics among the connected BiGRU layers and thus avoids the representation smoothness caused by the deep structure~\cite{he2016deep}.
In particular, we fuse each level representation of the video and sentence, i.e., $\mathbf{V}^x=[\mathbf{v}_{1}^{x},\mathbf{v}_{2}^{x},\cdots,\mathbf{v}_{T}^{x}]$ and $\tilde{\mathbf{s}}^x$, $x \in \{g,a,o\}$, and feed them into the Res-BiGRUs as follows,

\begin{equation}
\label{eq6}
\begin{aligned}
\qquad \quad \hat{\mathbf{F}}^x =&\text{Res-BiGRUs}^x(\mathbf{F}_{0}^x): \\
&\begin{cases}
  \mathbf{F}^x_0&=[\mathbf{v}_{1}^{x}\| \tilde{\mathbf{s}}^x,\mathbf{v}_{2}^{x}\| \tilde{\mathbf{s}}^x,\cdots,\mathbf{v}_{T}^{x}\| \tilde{\mathbf{s}}^x]\\
  \mathbf{H}^{x}_m &=f^x_m(BiGRU_m^x(\mathbf{F}_{m-1}^x)), \\
  \mathbf{F}^x_m &=ReLU(\mathbf{H}^{x}_m+\mathbf{F}^{x}_{m-1}), m={1,2,\cdots, M}
\end{cases}
\end{aligned}
\end{equation}
where $\text{Res-BiGRUs}^x$ denotes the Res-BiGRUs for the $x$ level representation. $BiGRU_m^x$ and $f^x_m$, $x \in \{g,a,o\}, m \in \{1,2,\cdots, M\}$ stand for the $m$-th BiGRU layer and the $m$-th fully-connected layer in the Res-BiGRUs of the fusion network $\mathcal{F}^x$, respectively.  $\mathbf{H}^{x}_m$ and $\mathbf{F}^x_m$ denote the hidden representation and output of the $m$-th BiGRU layer. Finally, we obtain the fused representation of the video and sentence as $\hat{\mathbf{F}}^x=\mathbf{F}_{M}^x  \in \mathbb{R}^{L \times d_f}$, where $d_f$ is the dimension of the fused representation. 

\subsubsection{Hierarchical Moment Localization}
We conduct the hierarchical moment localization based on the multi-level fused video and sentence representation $\hat{\mathbf{F}}^x, x \in \{g,a,o \}$. In particular, for each level, we first localize the coarse moment candidates of various lengths by the convolutional neural network. Then we refine the boundary of the moment candidates by predict their offsets to adjust the boundary. 

\indent \textbf{Candidate Ranking. }
Following previous studies \cite{gao2017tall, anne2017localizing, liu2018attentive, chen2018temporally, chen2019semantic, xu2019multilevel, zhang2019man, ge2019mac}, we employ the 1D-Convolutional neural network, which is able to jointly generate a series of fixed-length (depending on the filter size) moment candidates as well as their ranking scores with the 1-D convoluntion operation. In particular, we introduce $3$ 1D-CNN based ranking networks to derive the moment candidate list based on their global, action, and object level representations, respectively. Formally, we have
\begin{equation}
  \{(t_k^{x,s}, t_k^{x,e}, r_k^{x})\}_{k=1}^K = sigmoid(Rank^x(\hat{\mathbf{F}}^x)),
  \label{eq7}
\end{equation}
where $Rank^x, x \in \{g,a,o \}$ is the 1D-CNN based ranking network. For all the three ranking networks, we use $Q$ filters with the same set of sizes $\{w_1, w_2, \cdots, w_Q\}$, where $w_q$ is the size the $q$-th filter, based on which we can obtain a set of moment candidates of the size $w_q$. Since the $Q$ filters of all the ranking networks share the same set of sizes, the set of moments candidates of different lengths, denoted by the start and end points $(t_k^{x,s}, t_k^{x,e})$'s, derived from the representations of three different levels are the same. Namely, $(t_k^{g,s}, t_k^{g,e})=(t_k^{a,s}, t_k^{a,e})=(t_k^{o,s}, t_k^{o,e})$. For simplicity, we use $(t_k^{s}, t_k^{e})$ to denote the unified start and end points of $k$-th moment candidate. $r_k^{x}$ is the corresponding ranking score of the moment candidate $(t_k^{s}, t_k^{e})$ at the $x$ level. 

Ultimately, we learn the final ranking score $r_k$ of the $k$-th moment candidate by multiple levels as follows,
\begin{equation}
  r_k = f_h(r_k^g\| r_k^a \| r_k^o),
\end{equation}
where $f_h$ is a fully-connected layer.

\indent \textbf{Boundary Regression. }
To refine the moment localization, we further predict the offsets of candidate moments to adjust their boundaries. Similar with the ranking network, we also employ the 1D-Convolutional Neural Nework to predict the offsets for the candidate moments. It is worth noting that we only use the global representation to adjust the boundary, since it is more comprehensive compared with other representations.
\begin{equation}
  \left\{
  \begin{aligned}
  &\{d_k^{s}\}_{k=1}^K = Offset^s(\hat{\mathbf{F}}^g), \\
  &\{d_k^{e}\}_{k=1}^K = Offset^e(\hat{\mathbf{F}}^g),
  \end{aligned}
  \right.
\end{equation}
where $Offest^s$ and $Offest^e$ are the offset determination network regarding the start point and end point of the moment, respectively.  In particular, both $Offest^s$ and $Offest^e$ are implemented with the 1D-convolutional neural network. $d_k^{s}$ and $d_k^{e}$ refer to the start and end offset of the $k$-th moment candidate, respectively.

Ultimately, the predicted refined boundary $(\hat{\xi }_{k}^{s},\hat{\xi }_{k}^{e})$  of the $k$-th moment candidate can be represented as follows,
\begin{equation}
  \left\{
  \begin{aligned}
  &\hat{\xi }_{k}^{s} = t_{k}^{s} + d_{k}^{s}, \\
  &\hat{\xi }_{k}^{s} = t_{k}^{e} + d_{k}^{e}.
  \end{aligned}
  \right.
\end{equation}

\subsubsection{Training}
As for the optimization, we employ the alignment loss and regression loss to encourage the precious moment localization of the model.

\indent \textbf{Alignment Loss.}
Similar with~\cite{zhang2019man,chen2018temporally}, we use alignment loss to promote the model to assign a high ranking score to the candidate moment that has a large overlap with the ground truth. In particular, we use Intersection over Union (IoU) to represent the alignment degree between the candidate moment $(t_{k}^{s}, t_{k}^{e})$ and the ground truth moment $({\xi }_{k}^{s}, {\xi }_{k}^{e})$, which can be denoted as ${IoU}_k$. The alignment loss $\mathcal{L}_{aln}$ is formulated as, 
\begin{equation}
	\mathcal{L}_{aln} = -\frac{1}{K}\sum_{k=1}^{K} {IoU}_k\log(r_k) + (1-{IoU}_k)\log(1-r_k),
\end{equation}
where $K$ is the total number of the candidate moments. 

\indent \textbf{Regression Loss.}
To boost the boundary adjustment, we use the regression loss $\mathcal{L}_{reg}$ to promote the generated candidate moments close to the ground truth, which is formulated as follows, 

\begin{equation}
	\mathcal{L}_{reg} = smooth_{L_1}(\xi^s - \hat{\xi}_{k}^{s}) +smooth_{L_1}( \xi^e - \hat{\xi}_{k}^{e}),
\end{equation}
where $smooth_{L_1}$ denotes Smooth $L_1$ loss function. Pay attention that we only calculate the loss of the candidate moment $(\hat{\xi }_{k}^{s},\hat{\xi }_{k}^{e})$ with the specific $k$, which has the highest IoU with the ground truth moment $(\xi^{s},\xi^{e})$. 
Here we obtain the final loss function, which is defined as follows,
\begin{equation}
	\mathcal{L} = \mathcal{L}_{aln} + \alpha\mathcal{L}_{reg},
\end{equation}
where $\alpha$ is a trade-off hyper-parameter. 

\section{Experiments}
\label{Experiments}

\subsection{Datasets}
To evaluate our proposed HDRR model, we conducted experiments on two benchmark datasets. 

\indent \textbf{Charades-STA} \cite{gao2017tall}:
This dataset is built based on the Charades dataset \cite{sigurdsson2016hollywood}, which contains $6,672$ videos of indoor activities and $16,128$ query-video pairs. There are $12,408$ pairs for training and $3,720$ for testing. The average duration of each video is $29.76$ seconds. Each video has 2.4 annotated moments and each annotated moment lasts for $8$ seconds on average. 

\indent \textbf{ActivityNet-Captions} \cite{krishna2017dense}: 
This dataset contains $20K$ videos with $100K$ queries, where $37,421$ query-video pairs are used for training and $34,536$ for testing. The average duration of the videos is 110 seconds. On average, each video has $3.65$ annotated moments and each annotated moment lasts for $36$ seconds.

\begin{table}[t]
  \centering
  \small
  \caption{Performance comparison on Charades-STA dataset in terms of R@1, IoU@0.3, R@1, IoU@0.5 and R@1, IoU@0.7. “-” indicates that the  corresponding result is unavailable. }
  \vspace{-0.4cm}
  \label{ComparisionSTA}
  \begin{tabularx}
    {\linewidth}
    {X<{\centering}|X<{\centering}|X<{\centering}|X<{\centering}|X<{\centering}}
    \toprule
    \multicolumn{2}{c|}{} & \multicolumn{3}{c}{Charades-STA}  \\
    \hline
    Feature & Method & R@1, IoU@0.3 & R@1, IoU@0.5 & R@1, IoU@0.7 \\
    \hline
    \multirow{15}*{C3D} 
    & MCN & - & 17.46 & 8.01 \\
    & CTRL& -  & 23.63 & 8.89  \\
    & ABLR & - & 24.36 & 9.01  \\
    & SM-RL & - & 24.36 & 11.17 \\
    & SAP & - & 27.42 & 13.36\\
    & ACL & - & 29.39 & 12.23 \\
    & QSPN& 54.70  & 35.60 & 15.80  \\ 
    & DEBUG &54.95 & 37.39 & 17.69 \\
    & RWM& -  & 36.70 & 13.74 \\
    & CBP & 54.30 & 36.80 & 18.87 \\
    & GDP & 54.54 & 39.47 & 18.49 \\
    & TripNet & - & 38.29 & 16.07 \\
    & TSP-PRL & - & 37.39 & 17.69\\
    & PMI & 55.48 & 39.73 & 19.27 \\
    & \textbf{HDRR} & \textbf{62.37} & \textbf{43.04} & \textbf{21.32}  \\
    \hline
    \multirow{3}*{Twostream} 
    & RWM & - & 37.23 & 17.72 \\
    & TSP-PRL& -  & 45.30 & 24.73 \\
    & \textbf{HDRR} & \textbf{68.33} & \textbf{54.06} & \textbf{27.31}\\
    \hline
    \multirow{5}*{I3D} 
    & MAN& -  & 46.63 & 22.72 \\
    & ExCL& 65.10  & 44.10 & 23.30\\
    & SCDM & - & 54.44 & 33.43\\
    & LGI & 72.96 & 59.46 & \textbf{35.48}\\
    & \textbf{HDRR}& \textbf{73.44}  & \textbf{59.46} & 34.11\\
    \bottomrule
  \end{tabularx}
  \vspace{-0.6cm}
\end{table}

\begin{table}[h]
  \centering
  \small
  \caption{Performance comparison on ActivityNet-Captions dataset in terms of R@1, IoU@0.3 and R@1, IoU@0.5. “-” indicates that the corresponding results are not available.}
  \vspace{-0.4cm}
  \label{ComparisionActivity}
  \begin{tabularx}
    {\linewidth}
    {X<{\centering}|X<{\centering}|X<{\centering}|X<{\centering}}
    \toprule
    \multicolumn{2}{c|}{}  & \multicolumn{2}{c}{ActivityNet-Captions} \\
    \hline
    Feature & Method  & R@1, IoU@0.3 & R@1, IoU@0.5 \\
    \hline
    \multirow{17}*{C3D} 
    & MCN & 21.37 & 9.58 \\
    & CTRL & 28.70 & 14.00 \\
    & ACRN &31.29 & 16.17 \\
    & TGN & 43.81 & 27.93 \\
    & QSPN & 45.30 & 27.70 \\
    & TripNet & 48.42 & 32.19 \\ 
    & ABLR & 55.67 & 36.79 \\
    & RWM & 53.00 & 36.90 \\
    & CBP & 54.30 & 35.76 \\
    & SCDM & 54.80 & 36.75 \\
    & GDP & 56.17 & 39.27 \\
    & DEBUG &55.91 & 39.72 \\
    & TSP-PRL & 56.08 & 38.76 \\
    & LGI & 58.52 & 41.51 \\
    & 2D-TAN & 59.45 & \textbf{44.51} \\
    & PMI & 59.69 & 38.28 \\
    & \textbf{HDRR} & \textbf{61.10} & 43.20 \\
    \bottomrule
  \end{tabularx}
  \vspace{-0.6cm}
\end{table}

\subsection{Implementation Details}
\ \indent \textbf{Evaluation Metric.}
Similar with \cite{gao2017tall}, we adopted the metric ``R@$m$, IoU@$n$'', which represents the proportion of the top $m$ moment candidates with IoU larger than $n$, for evaluation. Following the mainstream test setting, we set $m$ as 1 for both datasets, while $n$ as {0.3, 0.5, 0.7} for Charades-STA dataset, and {0.5, 0.7} for ActivityNet-Captions dataset. 

\indent \textbf{Video Feature. }
As for the Charades-STA dataset, we used I3D network~\cite{carreira2017quo}, C3D network~\cite{tran2015learning}, and Twostream network~\cite{wang2016temporal} for the feature extraction. The extracted dimensions are $1,024$, $4,096$, and $8,192$ respectively. As for the ActivityNet-Captions dataset, we used the widely used features of $500$-D extracted by C3D network. As the average duration of the video in the ActivityNet-Captions dataset is longer than that in the Charades-STA dataset, we split each video in the Charades-STA and ActivityNet-Captions datasets into $75$ and $200$ units, respectively.

\indent \textbf{Text Feature.}
For the sentence presentation, we obtained the word embedding of $300$-D by the pre-trained GloVe \cite{pennington2014glove}.
We used the pre-trained BERT \cite{Shi2019SimpleBM} to conduct the semantic role labeling. The maximum length of the sentence in Charades-STA and ActivityNet-Captions datasets are set as $10$ and $50$, respectively. 

\indent \textbf{Training settings. }
A single NVIDIA Titan XP and a single NVIDIA RTX 2080Ti are used to train our model. 
We used Adam optimizer and set the learning rate as $0.003$ for Charades-STA dataset and $0.0003$ for ActivityNet-Captions dataset. The batch size is set to $128$ in both datasets. 
The depth $M$ of the Res-BiGRUs is set as 3. 
Similar with the existing work~\cite{qu2020fine}, the filter size in the 1D-convolutional neural network (i.e., $Rank^x$, $Offset^s$, and $Offset^e$ ) is set as [6, 12, 24, 48, 72] for Charades-STA dataset and [16, 32, 64, 96, 128, 160, 192] for ActivityNet-Captions dataset. 
The trade-off parameter $\alpha$ used for balancing the two losses is set as $0.001$. 

\subsection{Performance Comparison}
We compared our HDRR with the following state-of-the-art methods, including proposal-based methods, proposal-free methods, and reinforcement-learning-based methods. For proposal-based methods that generate dense proposals to localize the moment and perform location regression to adjust the boundary, we adopt CTRL \cite{gao2017tall}, MCN \cite{anne2017localizing}, ACRN \cite{liu2018attentive}, TGN \cite{chen2018temporally}, SAP \cite{chen2019semantic}, QSPN \cite{xu2019multilevel}, MAN \cite{zhang2019man}, ACL \cite{ge2019mac}, as well as the latest work CBP \cite{wang2020temporally}, 2D-TAN\cite{zhang2020learning}, SCDM \cite{yuan2019semantic}, and PMI \cite{chen2020learning} as the baselines. Regarding the {proposal-free methods} that predict the results directly without the candidate boxes, we choose ABLR \cite{yuan2019find}, ExCL \cite{ghosh2019excl}, DEBUG \cite{lu2019debug}, as well as the latest work GDP \cite{chen2020rethinking}, and LGI \cite{mun2020local} as the baselines. Pertaining to the {Reinforcement-learning based methods}, where the reinforcement learning is used, we select RWM \cite{he2019read}, SM-RL \cite{wang2019language}, TripNet \cite{hahn2019tripping}, as well as some latest work TSP-PRL \cite{wu2020tree} as the baseline. 

Tables \ref{ComparisionSTA} and \ref{ComparisionActivity} show the performance comparison among our HDRR and the state-of-the-art methods on two datasets. The experiment results of the baseline methods are referred by their papers. From these two Tables, we have the following observations:
1) Our HDRR has superiority over other methods among most scenarios, demonstrating the effectiveness and robustness of our model.
2) Our HDRR largely surpasses the baseline methods on the Charades-STA dataset with the two-stream feature. 
For instance, HDRR surpasses TSP-PRL by the margin of $8.76\%$ with respect to R@1, IoU@0.5. 
The possible reason may be that the two-stream feature is more powerful in capturing the actions and objects information of the video, which hence boosts the multi-level video representation learning in Section \ref{Hierarchical Video Encoder} and thus improves the localization performance. 

\begin{table}[t]
  \centering
  \small
  \caption{Results of the ablation studies on the Charades-STA dataset based on the I3D features.}
  \vspace{-0.4cm}
  \label{AblationStudy}
  \begin{tabular}{c|c|c|c}
    \toprule
    \multirow{2}{*}{Method} 
    & R@1, & R@1, & R@1, \\
    & IoU@0.3 & IoU@0.5 & IoU@0.7 \\
    \hline
    HDRR-w/o-Action & 70.00  & 57.23 & 31.77 \\
    HDRR-w/o-Object & 71.99 & 58.80 & 31.72  \\
    HDRR-w/o-Action\&Object & 69.76 & 55.27 & 30.13 \\
    HDRR-w/o-Res-BiGRUs & 68.60  & 51.34 & 25.97   \\  
    HDRR-w/o-self-attention & 72.45 & 58.33 & 32.77  \\
    \hline
    \textbf{HDRR} & \textbf{73.44} & \textbf{59.46} & \textbf{34.11} \\
    \bottomrule
  \end{tabular}
  \vspace{-0.6cm}
\end{table}
\subsection{Ablation Study}
We conducted the ablation study to demonstrate the effects of different components of our HDRR with the following derivations: 
\begin{itemize}
  \item \textbf{HDRR-w/o-Action}: We removed the action  level representation of the video and sentence to verify the effect of the action level representation. 
  
  \item \textbf{HDRR-w/o-Object}: We removed the object level representation of the video and sentence to investigate the effect of the object  level representation.
  
  \item \textbf{HDRR-w/o-Action\&Object}: We removed both the action and object level representations of video and sentence.  
  
  \item \textbf{HDRR-w/o-Res-BiGRUs}: To evaluate the effectiveness of the proposed Res-BiGRU module, we replaced the Res-BiGRUs with a fully-connected layer in HDRR.
  
  \item \textbf{HDRR-w/o-Residual}: We removed the residual connections in Res-BiGRUs to test the effect of the residual incorporation. 
\end{itemize}

Table \ref{AblationStudy} presents the results of the ablation study, from which we can draw the following conclusions: 
1) Our HDRR surpasses HDRR-w/o-Action, HDRR-w/o-Object, and HDRR-w/o-Action\&Object, which indicates that both of the action and object level representations are able to provide useful information to boost the moment localization.
2) Our HDRR achieves better performance than HDRR-w/o-Res-BiGRUs, demonstrating the effectiveness of the designed Res-BiGRUs. The deep and residual structure of the Res-BiGRUs is able to facilitate the adaptive integration between the video and sentence, and enhance the moment localization. 
And 3) our HDRR surpasses HDRR-w/o-self-attention, demonstrating the effectiveness of the multi-head self-attention in capturing the prominent information of the sentence and enhancing the understanding of it.

\subsection{Result Visualization}
\begin{figure}[t]
  \centering
  \includegraphics[width=1\linewidth]{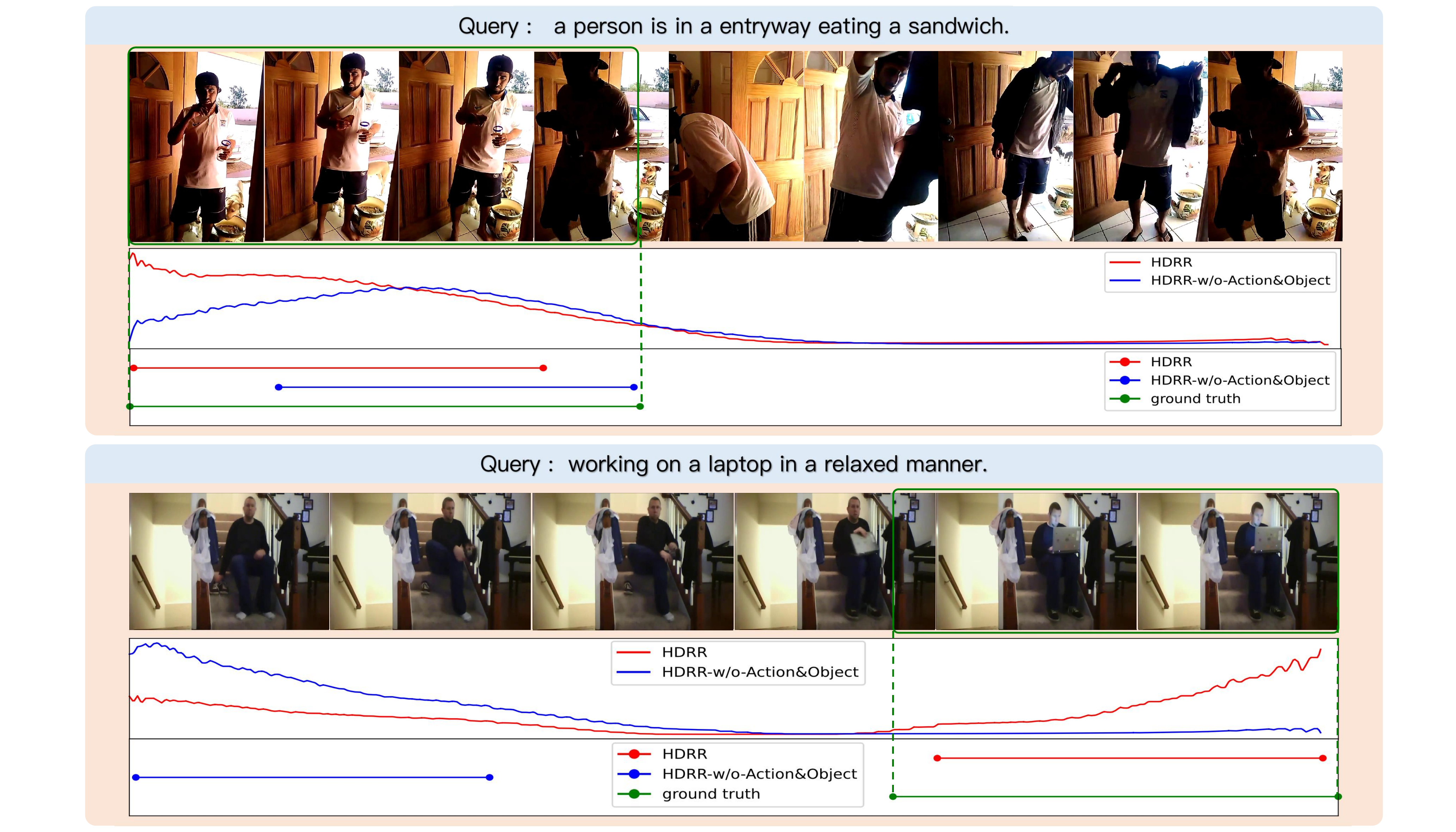}
  \vspace{-0.6cm}
  \caption{Visualization of the ranking score distributions as well as the localization results of the query-video pairs.} 
  \Description{A picture that visualizes our results. }
  \label{VisualizationHierarchical}
  \vspace{-0.5cm}
\end{figure}

To get an intuitive understanding of our model, we illustrated the ranking score distributions and the final localization of the query-video pairs, with two examples of HDRR and HDRR-w/o-Action\&Object in Figure \ref{VisualizationHierarchical}. 
In the first example, with the enhancement to the representation of the action ``eating'' and the object ``sandwich'', our HDRR pays more attention to the beginning of the video and localizes the moment closer to the ground truth. In the second example, we found that both the ranking score distribution of the beginning and the end candidates are higher, which may caused by the content similarity of the beginning and the end of the video. Owing to the attention on the object information (i.e., ``laptop''), our HDRR achieves more reasonable localization results than HDRR-w/o-Action\&Object. These two examples demonstrate the effectiveness of the hierarchical moment localization based on the multi-level representations of our HDRR. 

\section{Conclusion}
\label{Conclusion}
In this work, we propose a Hierarchical Deep Residual Reasoning (HDRR) model to solve the problem of temporal moment localization, where the hierarchical matching is conducted on the video and sentence based on their multi-level representations. 
In addition, we design the simple yet effective Res-BiGRUs for feature fusion, which is able to adaptively grasp useful information in exploring deeper understanding of the two modalities, alleviating the problem of model design caused by different difficulty of data understanding.
Through the methods above, the two modalities achieve finer-grained exploration and interaction. 
Extensive experiments conducted on the Charades-STA and ActivityNet-Captions datasets demonstrate the promising performance of our HDRR. 
In the future, we will dedicate to the investigation of more appropriate action and object extraction manner in the video, to match the multi-level text representations.

\begin{acks}
\label{Acknowledgments}
This work is supported by the National Natural Science Foundation of China, No.:U1936203; the Shandong Provincial Natural Science Foundation, No.:ZR2019JQ23; CCF-Baidu Open Fund, No.: CCF-BAIDU OF2020019; Young creative team in universities of Shandong Province, No.:2020KJN012.
\end{acks}
 
\clearpage

\appendix

\end{document}